\documentclass[12pt]{article}
\newcommand{\degr}{$^{\circ}\,$}
\usepackage{graphicx,amssymb,amsmath}
\usepackage{cite}

\begin{document}
\centering{{\LARGE \bf Event Reconstruction for a DIRC}}\\
\vspace{ 1cm}
\begin{center}
\large{ V. Kyryanchuk $^a$ \thanks{on leave of absence from the Institute for Nuclear Research, National  Academy od Sciences of Ukraine, Kyiv, Ukraine} , H. Machner $^{a,b}$ \thanks{e-mail: h.machner@fz-juelich.de}, R. Siudak $^a$\thanks{permanent address: Institute of Nuclear Physics, PAN, Krakow, Poland} , J. Ritman$^a$}
\end{center}

%%\vspace{0.5 cm}
\begin{center}
{{$^a$}Institut f\"{u}r Kernphysik and J\"{u}lich Centre for Hadron Physics, Forschungszentrum J\"{u}lich, 52425~J\"{u}lich, Germany \\
{$^b$}Fachbereich Physik, University Duisburg-Essen,  Lotharstr. 1, 47048~Duisburg, Germany}

\vspace{1cm}

\small{
{\bf Abstract}

Monte Carlo simulations were performed to investigate  the possibility to add a DIRC detector to the WASA detector at the COSY accelerator. A statistical method for pattern recognition is presented and the possible angle resolution and velocity precision achieved are discussed.}

\end{center}

%%\keywords{WASA; DIRC; Multianode photomultiplier tubes; Reconstruction method}
%%\maketitle

\section{Introduction}
It is planned to measure rare decays of $\eta'$ mesons with the WASA at COSY detector \cite{WASA-COSY}, \cite{Adolph09}. The foreseen production reaction is
 \begin{equation}
 p+p\to \eta'+p+p
 \end{equation}
where the two protons will be measured and the $\eta'$ will be identified via missing mass technique. The cross section for this process is small, because no nucleon resonance couples strongly to the $\eta' p$ channel, in contrary to $\eta$ production. The $\eta'$ peak will be on top of a huge multi-pion background. The only chance to identify the $\eta'$ is to measure the four momentum vectors of the two protons with high accuracy. Although their emission angles $(\theta,\phi)$ are measured with sufficient precision, their kinetic energies are poorly determined by the present setup. A possible option to achieve sufficient resolutions seems to be a DIRC detector (Detection [of] Internally Reflected
Cherenkov [light]) \cite{Coyle94} in the forward detection system of WASA.

Cherenkov detectors are widely used in high energy physics. They can have different applications: particle identification, energy measuring, threshold detector for detecting specific type of particles in a definite energy range.

Our aim is to measure the velocity of protons in the kinetic energy range from $T=400$ MeV to $T=1000$ MeV with a precision better than $\Delta\beta/\beta=0.5\%$

\section{DIRC Design}
\begin{figure}[!h]
\begin{center}
\includegraphics[width=0.5\textwidth]{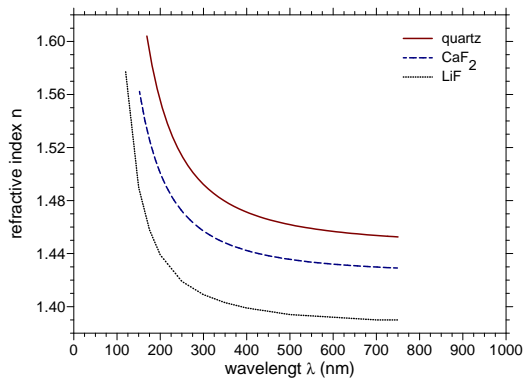}
\caption{The refractive index $n$ as function of the wavelength for the indicated materials (from Refs. \cite{Schott,Korth}). }
\label{Fig:refr}
\end{center}
\end{figure}

\begin{figure}[!h]
\begin{center}
\includegraphics[width=0.5\textwidth]{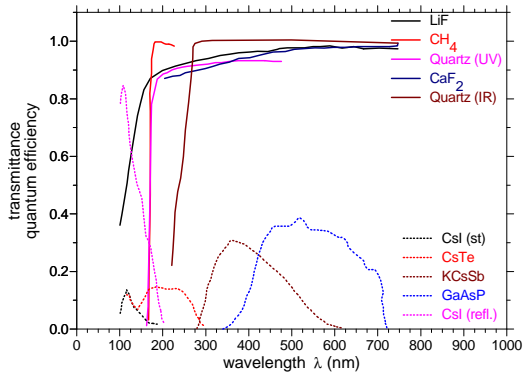}
\caption{Radiator transmittances (solid curves)\cite{Schott} and quantum efficiencies of photocathodes (dotted curves)\cite{Carruthers, Breskin, Kaufmann, Braema03} as function of the wave length. The bulk transmittances are given for 10 mm thickness and reflection losses at the surface result in an additional reduction of the transmittance by 10$\%$.}
\label{Fig_1}
\end{center}
\end{figure}

With the Monte Carlo package GEANT3 \cite{GEANT} different radiator materials were studied: LiF, CaF$_2$, quartz (fused silica) and plexiglas. All materials have different refractive indices which are smooth functions of the wave length in the visible range (see Fig. \ref{Fig:refr}). Quartz has the largest values for the refraction index $n$ at any given wavelength $\lambda$ compared to the two other materials. $n$ varies from 1.60 in the UV range to 1.45 in the IR range. The transmittances are rather independent from the wavelength above a certain ``threshold'' wavelength as is shown in Fig. \ref{Fig_1} (data taken from Ref. \cite{Schott} and \cite{Korth}). The radiator should have a large refractive index and high
transmittance to measure particles with ``low'' kinetic energy. LiF has higher transmittance in the UV range compared to CaF$_2$ and quartz. This gives the possibility to build a Cherenkov detector with LiF as radiator and a proper photocathode for detecting photons in the UV range. However, in the visible range, where the transmittance of LiF and quartz are compatible, quartz has the larger refractive index. We therefore omit CaF$_2$ and LiF in the further studies.

The quantum efficiencies for different photocathodes are also shown in Fig. \ref{Fig_1} (from Refs. \cite{Carruthers, Breskin, Kaufmann, Braema03}). CsI and CsTe are used for photon detecting in the UV range. Bi-alkali (KCsSb) and multi-alkali (GaAsP) photocathodes are used for the near UV and the visible light range. These types of photocathodes have larger and broader
distributions of the corresponding quantum efficiency function. This could compensate a lower number of emitted photons in this range of wavelengths. In a first step, simulations were performed for a RICH type Cherenkov detector with different radiators and photon detectors. However, a DIRC type detector was found to be favorable due to following reasons:  energy losses in the detector, the broad range of velocities to be measured in the proximity to the Cherenkov threshold and the  availability of very fast photon detectors. Based on the above discussed properties of the materials we limit ourselves in the simulations of DIRC detectors to fused silica and UV transparent plexiglas as radiator and photomultipliers with bialkali photocathodes.

\begin{figure}[!h]
\begin{center}
\includegraphics[width=0.5\textwidth]{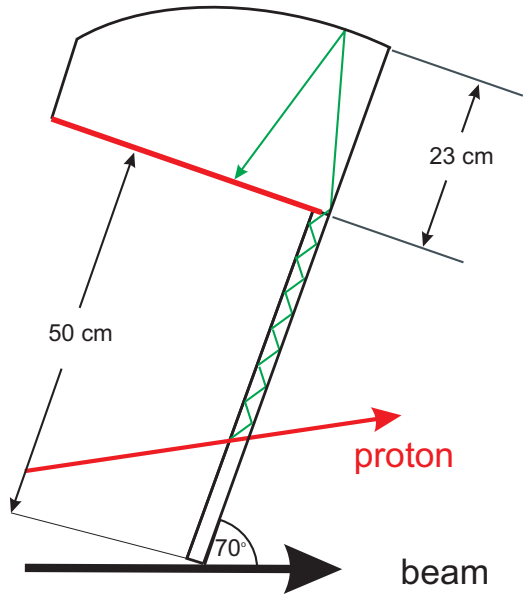}
\caption{Schematic side view of the detector. The radiator bars are inclined by 70 degree with respect to the  beam axis. A simulated proton is indicated with an arrow, the Cherenkov light as the zigzag and arrow. The focussing bar (mirror) is on top.}
\label{Fig_3}
\end{center}
\end{figure}

\begin{figure}[!h]
\begin{center}
\includegraphics[width=0.5\textwidth]{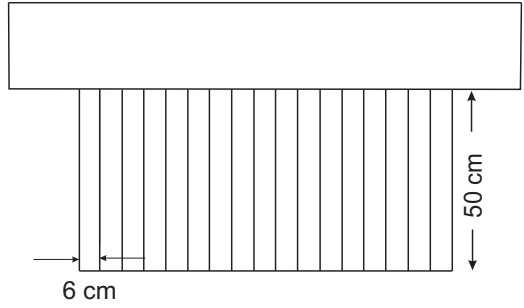}
\caption{Assembly of several bars to one module. Two such modules are foreseen, one above and one below the beam.}
\label{Fig_3a}
\end{center}
\end{figure}

We will now concentrate on the DIRC detector. The design differs from the Babar DIRC \cite{Coyle94, Cohen03}. It consists of radiator bars with dimensions 2x6x50 cm$^{3}$ and a focusing element with a cylindrical mirror (Figs. \ref{Fig_3} and \ref{Fig_3a}). The dimensions were for fused silica whereas for plexiglas an increased thickness of 4 cm was assumed.The individual modules are arranged in two planes, one above and one below the beam axis. The normal axis to each plane is tilted by 20\degr away from the beam axis (see Fig. \ref{Fig_3}). Emitted Cherenkov photons create an image on a pixelised screen. Each
pixel is a bi-alkali photocathode with dimension 5x5 mm$^{2}$.


\section{Monte Carlo Results}
A single event image consists of two arcs, which are the results of the internal reflection from the left and right side of the radiator bar. The shape of the arcs depend on the kinetic energy and particle species. The relative positions of the arcs depend on the horizontal angle of the detected particle track: the larger the horizontal angle, the larger the distance between arc centers \cite{Coyle94}.

\begin{figure}[!h]
\begin{center}
\includegraphics[width= 0.45\textwidth]{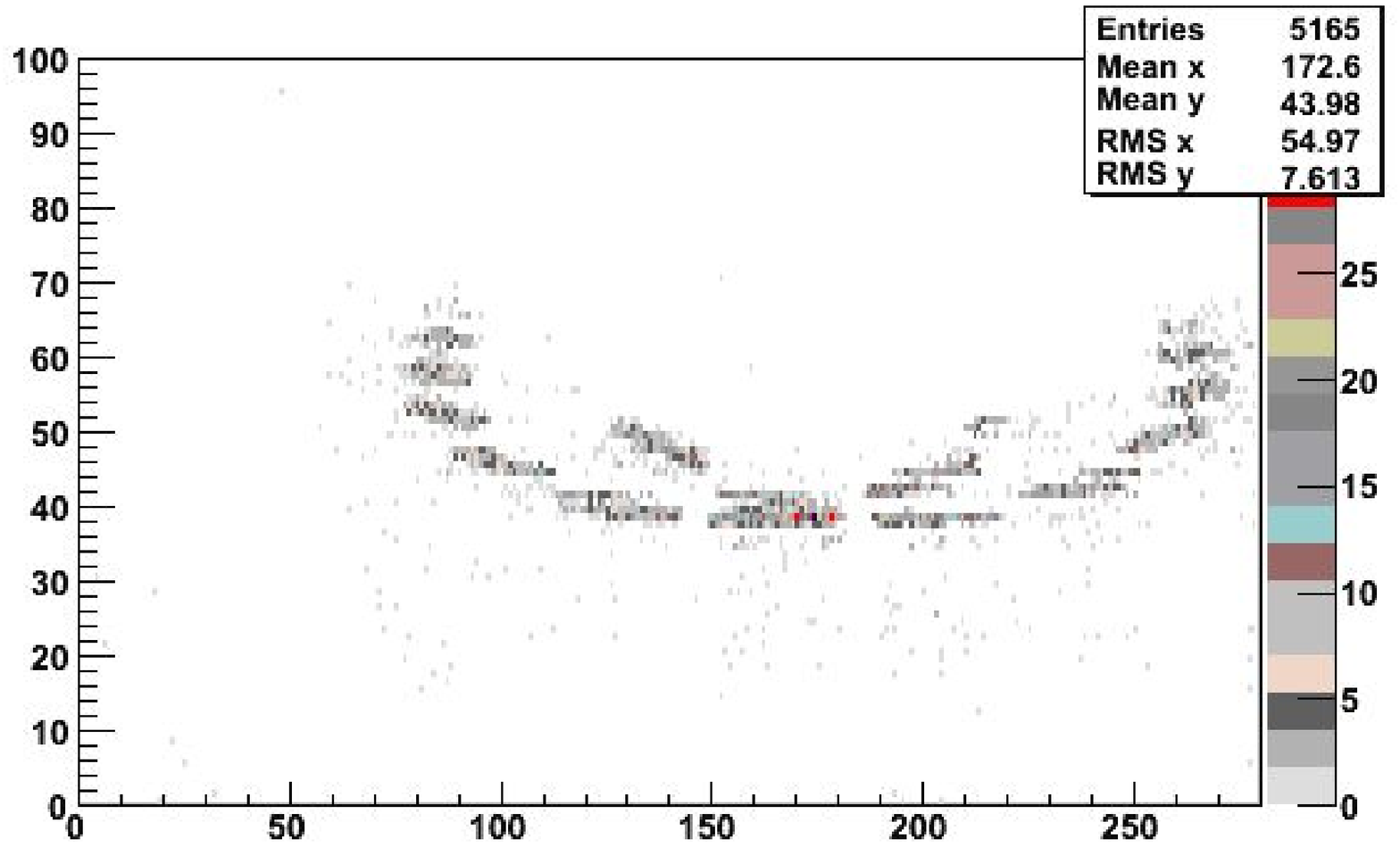}
\includegraphics[width=0.45\textwidth]{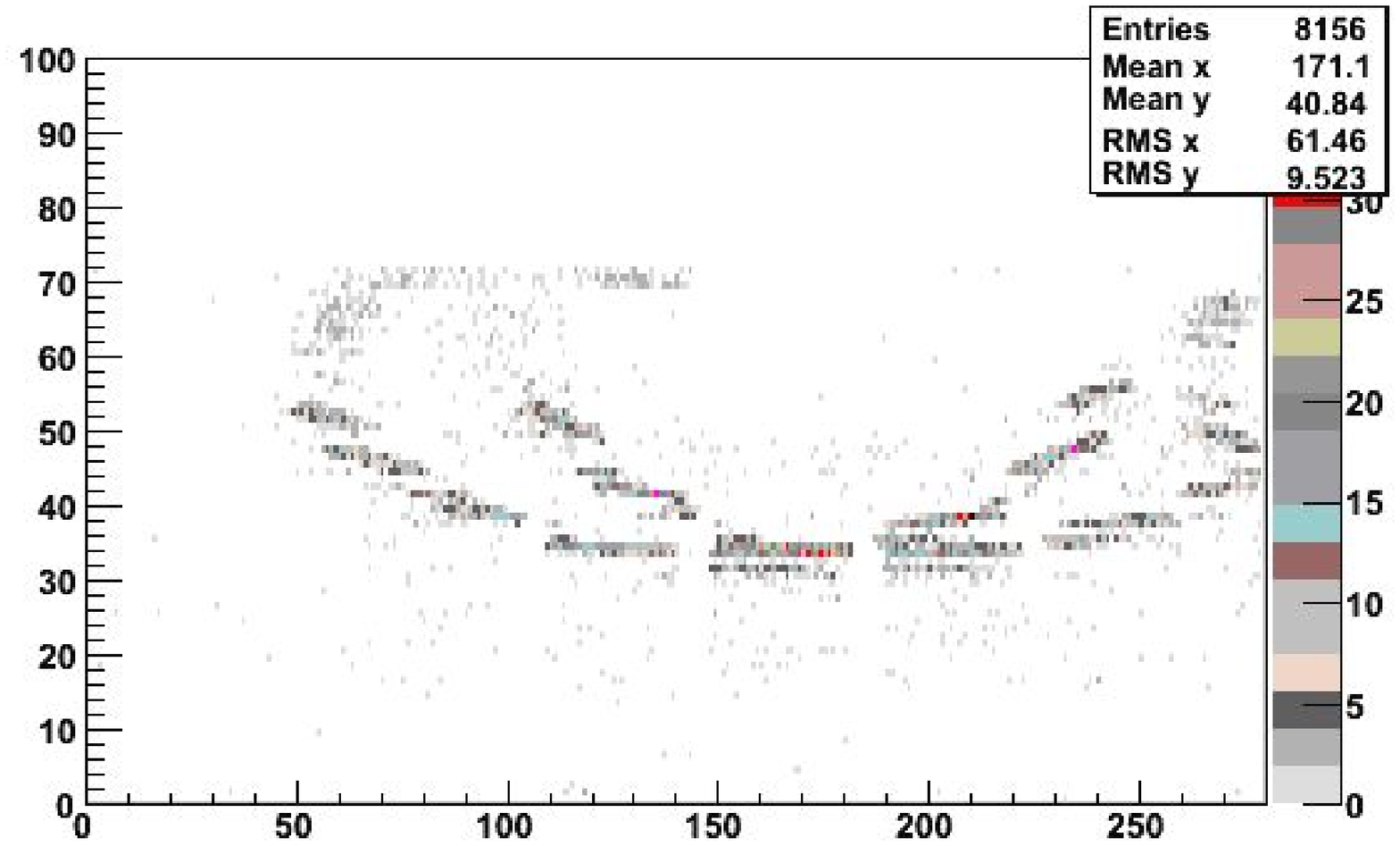}
\caption{Left: Scatter plot of the position of detected photons from 100 simulated events for protons with kinetic energy $T$= 600 MeV. The abscissa and ordinate are the horizontal and vertical position numbers on the photocathode. Right: Same as left but for $T$=1000 MeV}
\label{Fig_arcs}
\end{center}
\end{figure}

We have simulated Cherenkov images for protons with 600 MeV and 1000 MeV kinetic energy. We applied a left handed coordinate system with the $z$-axis in the beam direction,  $x-z$ defines the horizontal plane and $y$ is vertical to this plane.  The angle between the ejectile momentum and the (x-z) plane is $\theta_\text{vert}$ and the angle between the momentum projected in the plane and the beam direction is $\theta_\text{horiz}$. Thus $\theta_\text{horiz}=\arctan(z_x/p_z)$ and $\theta_\text{vert}=\arcsin(p_y/p)$.  Simulations were performed for track directions $\theta_\text{horiz} \sim 8.5$\degr fixed and $\theta_\text{vert}$ varied  and vice versa: $\theta_\text{vert} \sim 8.5$\degr fixed and $\theta_\text{horiz}$\ varied. They are shown in Fig. \ref{Fig_arcs}.

\begin{figure}[!h]
\begin{center}
\includegraphics[width=8 cm]{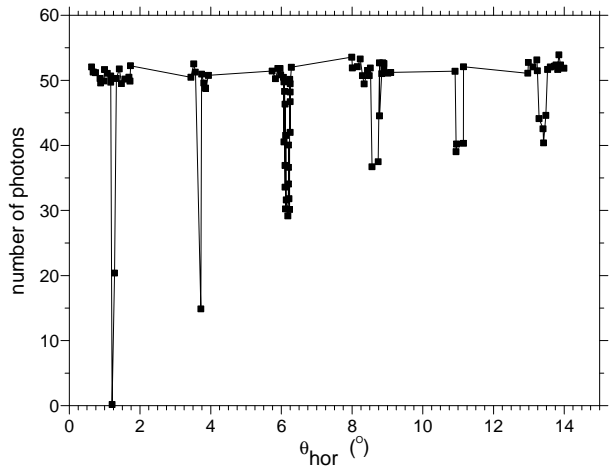}
\caption{The average number of the detected photons as function of the horizontal angle for protons with $T$ = 600 MeV. }
\label{Fig_9}
\end{center}
\end{figure}

The average number of emitted photons depends on the path through the radiator end the kinetic energy. Fig. \ref{Fig_9} shows the dependence of the number of detected photons on the horizontal track direction for protons with kinetic energy $T$ = 600 MeV. The number of detected photons drops sharply when the particle passes through the
edges of the radiator bars. This effect results from the detector
construction.

Furthermore, image distortion is also observed. Smeared arcs or even four arcs could be observed when the particle goes through the edges of two neighboring bars.

\section{Event Reconstruction}

The following method is used for event reconstruction. A collection of events with different particle species as well as energies and directions is simulated. Then a second set of simulated data named detected are ``compared'' to the simulated data. The comparison is realized by the following function $F$ (\ref{equ_1})
\begin{equation}\label{equ_1}
F(k,(E,P))=\sum\limits_{i,j} {det (i,j)\cdot \overline {sim(i,j)} }.
\end{equation}
Here $k$ denotes the particle species, $E,P$ the energy and momentum of the particle, $det(i,j)$ is the number of detected photons in $(i,j)$-th photoelement in the experiment and $\overline{sim(i,j)}$ the simulated average number of detected photons in $(i,j)$-th photoelement. The expression is a maximum when the experimental image on the screen is
similar to the simulated one. This method provides an opportunity to determine $\beta$ and the direction of a particle if the particle type is known. The particle species information has to come  from other sources.

\begin{figure}[!h]
\begin{center}
\includegraphics[width= 0.35\textwidth]{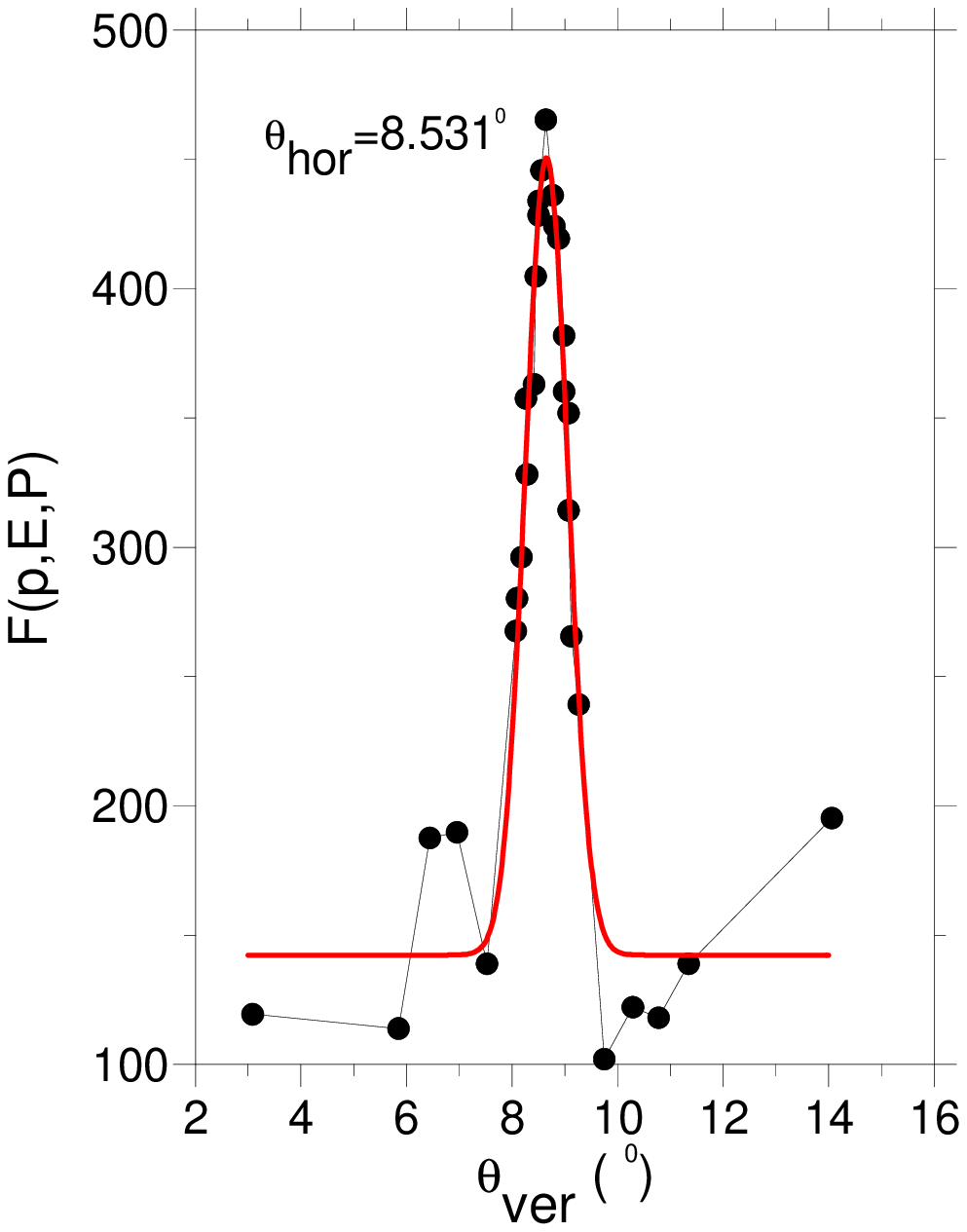}
\includegraphics[width= 0.35\textwidth]{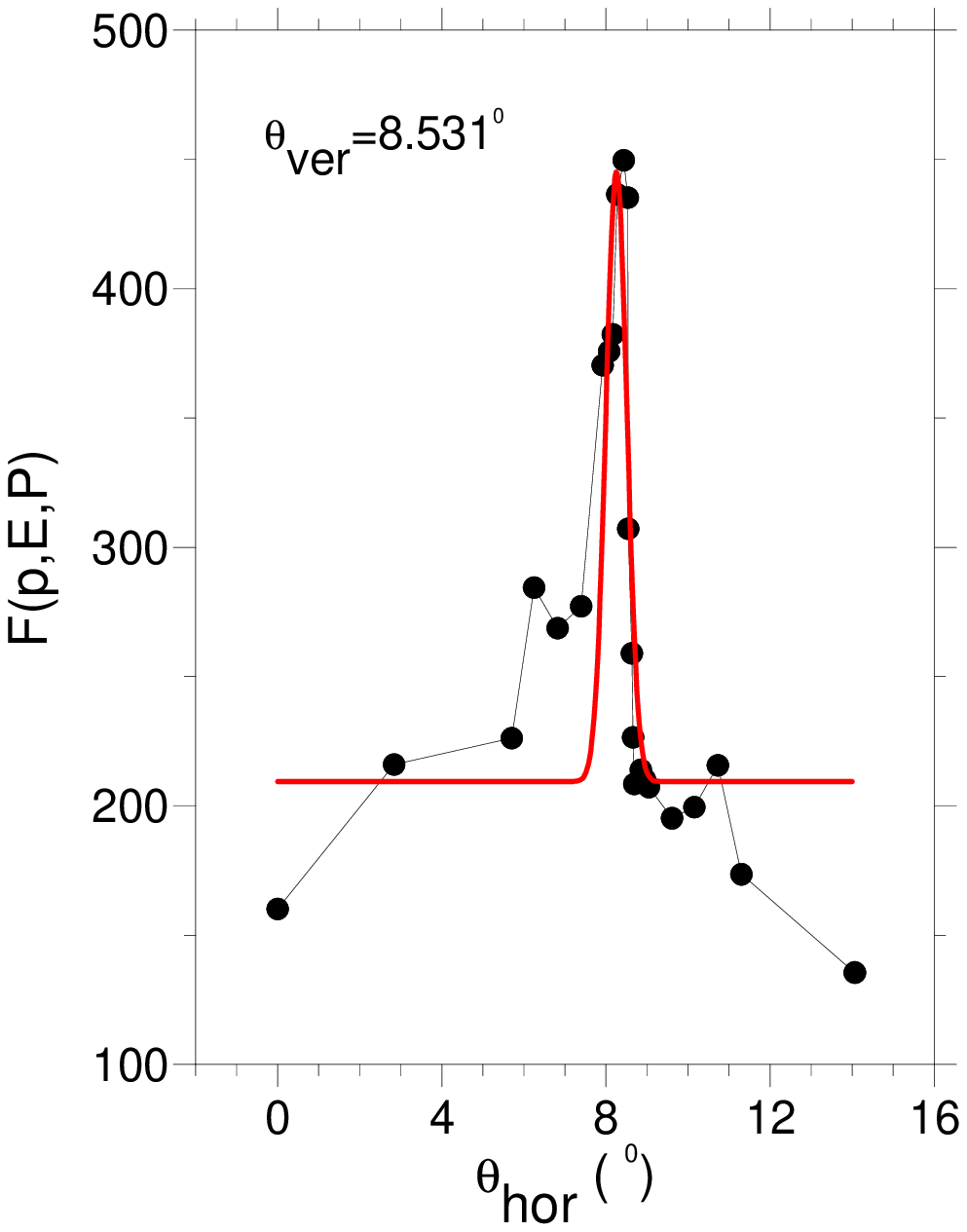}
\caption{Track reconstruction. Left: $F$ as function of the vertical angle and for fixed horizontal angle. The results (dots) are connected by lines to guide the eye. The smooth curve is a fit with a Gaussian plus a constant. Right: Same as left but horizontal and vertical angles are interchanged. }
\label{tab14}
\end{center}
\end{figure}

\begin{figure}[!h]
\begin{center}
\includegraphics[width=8 cm]{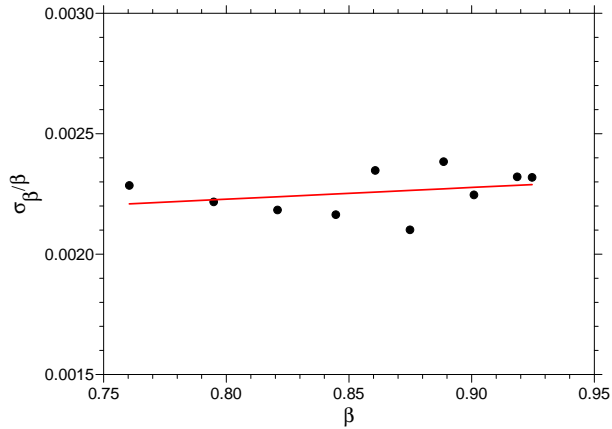}
\caption{Relative $\beta $-resolution for protons as function of $\beta$. The results are shown by dots, the solid line is a linear fit to the dots.}
\label{tab13}
\end{center}
\end{figure}

The possibility of track reconstruction for protons with $T$=600 MeV is demonstrated in Figs. \ref{tab14} and \ref{tab13}. The particle direction is extracted under the assumption that the particle species and its energy are known. Horizontal and vertical angles relative to the beam direction are reconstructed given that one of both is fixed. The small widths ($\sigma$) of Gaussians fitted to the data of $\sim $0.3\degr and $\sim $0.5\degr indicate the sensitivity of the method. Similar results were obtained for protons with $T$=1000 MeV.


Fig. \ref{tab13} shows the $\beta $-resolution for protons. Here we have assumed that the angle measurement is performed with other detectors. The method gives a resolution ($\sigma_\beta/\beta$) for protons better than 0.3{\%}.  The results obtained can be transformed into the missing mass resolution. For an incident proton beam with 3.3 GeV/c momentum a missing mass resolution of $\sigma_{mm}\approx 0.5$ MeV is achieved.

\section{Summary}
We have designed a DIRC detector to measure protons with kinetic energy from 400 to 1000 MeV. Monte Carlo simulations for such a device have been performed. A method is discussed to deduce four momentum vectors of the protons. The method gives reliable information for the event reconstruction: velocity-resolution for
protons is better than 0.3{\%}. The routine is able to reconstruct tracks and identify particles when additional information is
provided by complementary detectors. The large data base of simulated events is a disadvantage of this method; however, optimization of the algorithm will increase the speed.

\end{document}